\documentclass[12pt]{article}
\usepackage{amsfonts}
\usepackage{amssymb}
\usepackage{amsthm}
\usepackage{hyperref}
\usepackage{amssymb,amsmath,amsthm}
\usepackage{enumerate}

\newcommand{\sss}{\setcounter{equation}{0}}
\newtheorem{theorem}{THEOREM}[section]

\newtheorem{lemma}[theorem]{LEMMA}

\newtheorem{corollary}[theorem]{COROLLARY}

\newtheorem{prop}[theorem]{PROPOSITION}

\newtheorem{remark}[theorem]{REMARK}

\newcommand{\ere}{ {\mathbb R}}

\newcommand{\CE}{{\mathbb C}}

\def\beq{\begin{equation}}
\def\ene{\end{equation}}

\def \ds {\displaystyle}
\newcommand{\bull}{\hfill $\Box$}

\def\qed{\ifhmode\unskip\nobreak\fi\ifmmode\ifinner
\else\hskip5pt\fi\fi\hbox{\hskip5pt\vrule width4pt height6pt
depth1.5pt\hskip1pt}}
\def\1{L^{p+1}}
\def\p{L^{ \frac{p}{p+1}}}

\def\P{L(P)}
\def\m{\mathcal M}
\def\var{\varphi}
\newcommand{\Q}{{\mathbb H}_{Q}}

\begin{document}
\baselineskip=20 pt
\parskip 6 pt

\title{The Initial Value Problem, Scattering and Inverse Scattering, for  Non-Linear Schr\"odinger Equations with a
Potential and a Non-Local Non-Linearity
\thanks{ PACS classification scheme (2003): 03.65.Nk, 02.30.Zz, 02.30.Jr, 03.65.Db} \thanks{ Research partially
supported by
Universidad Nacional Aut\'onoma de M\'exico under Project
PAPIIT-DGAPA IN 105799, and by CONACYT under Project P42553­F.}}
 \author{ Mar\'{\i}a de los \'Angeles Sandoval Romero and Ricardo Weder\thanks{ †Fellow Sistema Nacional de Investigadores.}
\\Instituto de Investigaciones en Matem\'aticas Aplicadas y en Sistemas \\Universidad Nacional Aut\'onoma de
M\'exico \\
Apartado Postal 20-726, M\'exico DF 01000
\\weder@servidor.unam.mx}

\date{}
\maketitle
\begin{center}
\begin{minipage}{5.75in}
\centerline{{\bf Abstract}}
\bigskip
We consider non-linear Schr\"odinger equations with a potential,
and non-local non-linearities, that are models in mesoscopic
physics, for example of a quantum capacitor, and that also  are
models of molecular structure. We study in detail the initial
value problem for these equations. In particular, existence and
uniqueness of local and global solutions, continuous dependence on
the initial data and regularity. We allow for a large class of
unbounded potentials. We have no restriction on the
growth at infinity of the positive part of the potential.

We also construct the scattering operator in the case of potentials that go to zero at infinity. Furthermore, we give a
method for the unique reconstruction of the potential from the small amplitude limit of the scattering operator.
In the  case of the quantum capacitor, our method allows us to uniquely reconstruct all the physical parameters
from the small amplitude limit of the scattering operator.

\end{minipage}
\end{center}
\newpage
\section{Introduction}
\sss

In recent years there is a considerable interest in non-linear
Schr\"odinger equations with a potential (NLSP) and a non-local
non-linearity that is concentrated in a bounded region of space,
for example, to model physical situations that appear in
mesoscopic physics. In particular, in \cite{s1,s2,s3,s4,s5} the following equation was introduced to
model a quantum capacitor (in \cite{s3} more general equations are discussed).

\beq
i\frac{\partial}{\partial t} u(x,t)= -\frac{d^2}{d x^2}
u(x,t)+ V_0 \, u(x,t) + \lambda Q(u)\chi_{[b,c]}(x)\, u(x,t),
\quad u(x,0)=\varphi(x),
\label{1.1}
\ene
with, $x,t \in \ere$, and
where we have set Planck's constant $\hbar$ equal to one and the
mass equal to $1/2$. For any $O\in \ere, \chi_{O}$ is the characteristic
function of $O$. The external potential, $V_0$,
is a double barrier,
$$
V_0(x)= \beta \left[\chi_{[a,b]}(x)+ \chi_{[c,d]}(x)\right],\qquad
\beta > 0,
$$
where, $ a < b <c < d$. Furthermore,

$$
Q(u):= \int_b^c\, |u(x,t)|^2 \, dx
$$
is the a-dimensional electric charge trapped in the well $[b,c]$.

We briefly describe the physical aspects of the model given by (\ref{1.1}), following \cite{s4}.
As is well known \cite{alw}, the interaction between electrons can play a crucial role in the electrical
transport properties of mesoscopic systems. In the case considered in \cite{s1}, \cite{s4}
a cloud of electrons move in a double barrier heterostructure in which the well region confined between the two
potential barriers acts like a   quantum capacitor whose energy depends on the electron charge trapped inside it.
The localization of the interaction is justified by the existence of a resonant state that allows for a long sojourn
time of the electrons inside the well. This leads to an accumulation of electric charge inside the quantum capacitor.
A main feature of equation (\ref{1.1}) is that the non-linearity is concentrated only in the region where the resonant
state is localized, i.e., within the two barriers. In \cite{s4}, among other results, a one mode approximation was considered,
an adiabatic condition was introduced, and numerical simulations where performed.

In this paper we study the following generalization of (\ref{1.1}),

\beq
i \frac{\partial}{\partial t}u(x,t) = -\frac{d^2}{d x^2}u(x,t) + V_0(x) u(x,t)+ \lambda \,\left(V_1u,u\right)
\, V_2(x)\,u(x,t), \quad u(x,0)=\varphi(x),
\label{1.2}
\ene
$x,t \in \ere$. By $(\cdot,\cdot)$ we denote the $L^2$ scalar product. The external potential $V_0$ is real valued,
and $V_1(x), V_2(x)$ are, in general, complex valued. $\lambda \in \CE$ is a coupling constant.

In the particular case where $V_0$ is a  double well, $V_1=V_2$ is real valued, and $\lambda \in \ere$, this
equation was extensively studied as a model for molecular
structure \cite{m1,m2}, \cite{m3,m5}, \cite{m4,m6,m7} and the
references quoted there. In these papers, molecular localization,
the suppression of the beating effect by the non-linearity, and
the semi-classical limit were studied, among other problems.

Below, we study in a detailed way the initial value problem for
(\ref{1.2}). We prove existence and uniqueness of local solutions
in $L^2, \Q^1,$ and $\Q^2$, where $\Q^j, j=1,2$, are Sobolev
spaces, and $  \Q^2 \subset \Q^1 \subset L^2$ (see section 2 for
the definition of $\Q^j, j=1,2$). We also  prove the continuity of
the solution on the initial value and a regularity result that
tells us that if the initial value belongs to $\Q^1$, the solution
in $\Q^1$ can not blow-up before the solution in $L^2$ does. We
obtain also a similar result on regularity between $\Q^1$ and
$\Q^2$ solutions, when the initial data belongs to $\Q^2$.
Furthermore, we prove that if $V_1,V_2,$ are real valued and
$\lambda \in \ere$ the $L^2$ solutions are global, i.e., they exist
for all times $t \in \ere$. If moreover, $V_1=V_2$ we prove that
the $\Q^j, j=1,2$, solutions are global. We prove these results for
a large class of unbounded external potentials, $V_0$ (see section
2). In fact, we have no restriction on the growth at infinity on
the positive part of the potential $V_0$. These results prove that
(\ref{1.2}) forms a dynamical system by generating a continuous
local/global flow \cite{ka2}. In this sense, the spaces $L^2,\Q^1$
and $\Q^2$ are fundamental for the equation (\ref{1.2}).

In \cite{m5}, \cite{m7} the existence and uniqueness of global solutions in the Sobolev spaces $\mathbb H^j, j=1,2$,
was proven in the case where the external potential $V_0$ is bounded.

Then, we consider external potentials that decay at
infinity, and we construct the  scattering operator for (\ref{1.2})
with reference dynamics given by the self-adjoint realization in
$L^2$ of $-\frac{d^2}{d x^2}$. Furthermore, we study the inverse
scattering problem. We prove that the small amplitude limit of the
scattering operator allows us to uniquely reconstruct $V_0$ and
$\lambda$. In the particular case of the quantum capacitor,
 (\ref{1.1}), this gives us all the physical parameters.

There is a very extensive literature on the initial value problem
and in scattering for the non-linear Schr\"odinger equation without
a potential and local non-linearities. As general references see,
for example, \cite{nls1,ka2,nls2,nls3,nls4} and
\cite{nls5}. For the case of non-linearities concentrated in a
finite number of points see \cite{nls6,nls7,nls8,nls9}.

For the initial value problem  for the NLSP with a local non-linearity and a
potential that has bounded second derivative see \cite{ka2}.  For the forced problem
on the half-line see \cite{we3}, where for local solutions there is no restriction
on the growth of the positive part of the potential at infinity, and for global
solutions only mild restrictions that allow, for example, for
exponential growth.

For  direct and inverse scattering for the NLSP with local
non-linearities see \cite{we2}, \cite{we5} and \cite{we6}, and for
the case of the non-linear Klein-Gordon equation with a potential
and  local non-linearities see \cite{we4} and \cite{we7}.
For an expository review of these results see \cite{we8}.
For the forced NLSP on the half-line see \cite{we9}.

For center manifolds for the NLSP with  local non-linearities see \cite{we10}, and
for  the non-linear Schr\"odinger equation with a double-well potential and a cubic local non-linearity see
\cite{sach}.

 The paper is organized as follows. In section 2 we prove our results on the
initial value problem. In section 3 we
 construct the scattering operator (direct scattering) and we prove our results on  inverse scattering.

\section{The initial value problem}
\sss In this section we absorb the coupling constant $\lambda$
into $V_2$ and we  consider the initial value problem for the
following NLSP,

\beq
i \frac{\partial}{\partial t}u(x,t) = -\frac{d^2}{d
x^2}u(x,t) + V_0(x) u(x,t)+ \left(V_1u,u\right) \, V_2\,u(x,t),
\quad u(x,0)=\varphi(x).
\label{2.1}
\ene

By $W_{j,p}, j=1,2, \cdots, 1 \leq p \leq \infty,$ we denote the Sobolev
space \cite{ad} of all functions in $L^p$ such that all its
derivatives of order up to $j$ are functions in $L^p$.  By $\|
\cdot \|_{j,p}$ we denote the norm in $W_{j,p}$.  By $\mathbb H^j,
j=1,2,\cdots$, we denote the $L^2$ based Sobolev spaces, $\mathbb H^j:=
W_{j,2}$.

We suppose that the potential $V_0$ satisfies the following
assumption.

\noindent {\it Assumption A}

\noindent Assume that,
\begin{gather}
V_0= V_{0,1}+ V_{0,2},\,{\rm with}\, V_{0,j}\in L^1_{\rm loc}, V_{0,1}\geq 0,
\label{2.2}\\
 \sup_{x\in \ere}\int_x^{x+1}
|V_{0,2}(x)|\, dx < \infty.
\label{2.3}
\end{gather}

 We denote , $Q:=\sqrt{V_{0,1}}$ and by $D(Q)$ the domain in $L^2$ of the operator of multiplication
by $Q$.

Let us denote by $H_0$ the self-adjoint realization of $-\frac{d^2}{d x^2}$ in $L^2$ with domain $\mathbb H^2$.
Equation (\ref{2.3}) implies that $V_{0,2}$ is  a quadratic form bounded perturbation of $H_0$ with relative bound zero
(this is proven, for example, as in the proof of \cite[Proposition 2.1] {we3}), that is, for any
$\epsilon >0$ there is a constant $K_\epsilon$ such that,

\beq
| (V_{0,2} \var,\var )| \leq \epsilon \| \var' \|_{L^2}^2+
K_\epsilon \|\var\|_{L^2}^2.
\label{2.3b}
\ene

It follows that the quadratic form,
\beq
h(\varphi,\psi):=(\var',\psi')+(V_0\var,\psi), \,{\rm with \, domain},\,D(h):=  \mathbb H^1 \cap D(Q),
\label{2.5}
\ene
is closed and bounded from below. Let $H$ be the associated bounded-below, self-adjoint operator (see, \cite{ka1, rs2}).
Then,
\beq
D(H)=\left\{\var \in D(h) :-\frac{d^2}{d x^2}\var + V \var \in L^2 \right\}
\, {\rm and}\, H\var= -\frac{d^2}{d x^2}\var + V \var ,\, \var \in D(H).
\label{2.6}
\ene

Let us take $N >0$ such that $H+N >1$ and let us denote $ D := \cap_{m=-\infty}^{\infty} D((H+N)^m)$. For any
$s\in \ere$ let $\mathbb H_Q^s$ be the completion of $D$ in the norm, $\|(H+N)^{s/2}\|_{L^2}$. Notice that
$ \Q^0=L^2, \mathbb H_Q^1 = D(\sqrt{H+N})=D(h)$ and that $ \mathbb H_Q^2= D(H)$.

Observe that the following norm is equivalent to the norm of $\Q^1$,

$$
\max\left[ \|\var\|_{\mathbb H^1},\,\|Q\var\|_{L^2}\right].
$$
Furthermore,  if $V_{0,1}=0 $ the potential $V_0$ is a
 quadratic form bounded perturbation of $H_0$ with relative bound
zero, and then,  $\Q^1= \mathbb H^1$. In this case the $\Q^1$
solutions that we consider below are just $\mathbb H^1$ solutions.

As $\Q^2=D(H)$, the following norm is equivalent to the norm of $\Q^2$,

$$
\max \left[\|\var \|_{\Q^1},   \left\|\left(-\frac{d^2}{dx^2}+V_0\right)\var\right\|_{L^2}   \right].
$$

The paper \cite{ka3} gives sufficient conditions in $V_{0} $  that assure that $ \Q^2 \subset \mathbb H^2$.
Furthermore, if
\beq
\sup_{x\in \ere}\int_x^{x+1}
|V_{0}(x)|^2\, dx < \infty,
\label{2.19}
\ene
it follows from  (\ref{2.3b}), and as $\|V_0 \var\|_{L^2}^2=(V_0^2\,\var, \var)$, and $(\var',\var')
\leq \|H_0\var'\|^2+\|\var\|^2$, that  $V_0$ is relatively bounded with respect to $H_0$ with relative bound
zero, that is, for any $\epsilon >0$ there is a constant $K_\epsilon$ such that,

\beq
\| V_{0} \var \| \leq \epsilon \| H_0 \var \|_{L^2}+ K_\epsilon \|\var\|_{L^2}.
\label{2.19b}
\ene
In this case (see \cite{ka1}, \cite{rs2}) $\Q^2 = \mathbb H^2$ and the $\Q^2$ solutions that we consider below
are just $\mathbb H^2$ solutions.

It follows from the functional calculus of self-adjoint operators that $H$ is bounded from $\Q^s$ to $\Q^{s-2}$ and
that $e^{-itH}$ is a strongly-continuous unitary group on $\Q^s, s \in \ere$. Moreover, for any $\varphi \in \Q^s,
s\in \ere,
e^{-itH} \varphi \in C\left(\ere, \Q^s\right) \cap C^1\left(\ere, \Q^{s-2}\right)$ and,

$$
i \frac{\partial}{\partial t} e^{-it H} \var = H e^{-it H} \var =  e^{-it H} H \var, \quad \var \in \Q^s.
$$

We introduce some further notation that we use below. Let $I=:[0,T],$ if $0 < T < \infty$  and $I= [0,\infty )$,
if $T=\infty$. For any Banach space $\mathcal X$ we denote by $\mathcal X_R$ the closed ball in $\mathcal X$ with
centre zero and radius $R$. If $ T < \infty$ we denote by $C(I,\mathcal X)$  the Banach space of continuous
functions from $I$ into $\mathcal X$ and if $T=\infty$ we denote by $C_B(I, \mathcal X)$  the Banach space of continuous
and bounded functions from $I$ into $\mathcal X$. For $T < \infty$, we define, $\mathcal N:= C(I,L^2)$ and
$\mathcal N^j := C(I, \Q^j), j=1,2$. For functions $u(t,x)$ defined in $\ere^2$ we denote $u(t)$ for $u(t,\cdot)$.

We study the initial value problem (\ref{2.1}) for $ t\geq 0$, but changing $t$ into $-t$ and taking the
complex conjugate of the solution (time reversal) we also obtain the results for $t \leq 0$.

By a $L^2$ solution on $I$ to (\ref{2.1})
we mean a function  $u\in C(I,L^2)\cap
C^1(I, \Q^{-2})$  that satisfies (\ref{2.1}).

Multiplying both sides of (\ref{2.1}) (evaluated at $\tau$) by
$e^{-i(t-\tau) H}$ and integrating in $\tau$ from zero  to $t$ we
obtain that,

\beq
u(t)=e^{-itH} \varphi + \frac{1}{i}\, G F(u),\, {\rm with}\, F(u):= (V_1 u,u)\, V_2 u
\label{2.7}
\ene
and where,
\beq
 G u := \int_0^t \, e^{-i(t-\tau)H} u(\tau)\, d\tau.
\label{2.8}
\ene
Moreover, let $u \in C(I,L^2)$ be a solution to (\ref{2.7}). Then, it follows from (\ref{2.7}) that
$ u \in C^1(I, \Q^{-2})$. We prove that $u$ solves (\ref{2.1})  taking the derivative of
both sides of (\ref{2.7}). Hence, equations (\ref{2.1}) and (\ref{2.7}) are equivalent. We obtain our results below
solving the integral equation (\ref{2.7}).

In the next theorem we prove the existence of local solutions in $L^2$.

\begin{theorem}\label{th2.1}
Suppose that assumption A is satisfied and that $V_j \in L^\infty, j=1,2$. Then, for any $\varphi \in L^2$ there is $0 <  T < \infty$,
 such that (\ref{2.1}) has a unique solution, $u \in C(I,L^2)$ with, $u(0)=\var$. $T$  depends only on
$\|\varphi\|_{L^2}$.
\end{theorem}

\noindent {\it Proof:}
We define,
\beq
\mathcal C(u): e^{-itH} \var +\frac{1}{i} G F(u).
\label{2.9}
\ene
We will prove  that we can take $R$ large enough, and $T$ so small, that $\mathcal C$ is a contraction on
$\mathcal N_R$.

It follows from Schwarz inequality that for any $u,v \in \mathcal N_R,$

\beq
\|F(u)-F(v)\|_{\mathcal N} \leq C \,(\|u\|^2_{\mathcal N}+\|v\|^2_{\mathcal N})\, \| u-v\|_{\mathcal N}
\leq  C \,2\, R^2 \,\|u-v\|_{\mathcal N}.
\label{2.10}
\ene
Then, as $e^{-itH}$ is unitary on $L^2$,

\begin{gather}
\|\mathcal C(u)\|_{\mathcal N}\leq \left[\|\var\|_{L^2} + C \, T \|u\|_{\mathcal N}^3 \right]
\leq \left[\|\var\|_{L^2} + C \,T\, R^3 \right], \label{2.11}
\\
\|\mathcal C(u)-\mathcal C(v)\|_{\mathcal N}\leq C T (\|u\|^2_{\mathcal N}+\|v\|^2_{\mathcal N})\, \| u-v \|_{\mathcal N}
\leq C \,T\, 2\, R^2 \| u-v \|_{\mathcal N}.
\label{2.12}
\end{gather}

Then, we can take $R,T$ such that $\|\var\|_{L^2} + C \,T\, R^3 \leq R$ and, $d:=  C \,T\, 2\, R^2 < 1$, what makes
$\mathcal C$ a contraction on $\mathcal N_R$. By the contraction mapping theorem \cite{rs1} $\mathcal C$ as a unique
fixed point, $u$, in $ \mathcal N$ that is a solution to (\ref{2.7}).

Suppose that there is another solution $v \in \mathcal N$. By the argument above, we have that $v(t)=u(t)$
for $ t \in [0,T_0]$ for some $ T_0 \leq T $. By iterating this
argument we prove that $ v(t)=u(t), 0 \leq t \leq T$.

\bull

We now prove that the solution depends continuously on the initial data.
\begin{theorem}\label{th2.2}
Suppose that assumption A is satisfied and that $V_j \in L^\infty, j=1,2$.
 Then, the solution  $u \in C([0,T],L^2),0< T < \infty,$
to (\ref{2.1}) with $u(0)=\var$,  given by theorem \ref{2.1}, depends continuously on the initial value $\var$. In a precise way, let
$\var_n \to \var$  strongly in $ L^2$. Then, for $n$ large enough, the solutions $u_n \in  C([0,T],L^2)$
to (\ref{2.1}) with initial values $\var_n$ exist and $ u_n \to u $ in $C([0,T],L^2)$.
\end{theorem}

\noindent{\it Proof:} We first prove a local version of the theorem with $T$ replaced by a $T_0$ small enough .
We define,
\beq
\mathcal C_n(u):= e^{-it H}\var_n+ \frac{1}{i} G F(u), \quad u \in \mathcal N(T_0) :=C([0,T_0],L^2).
\label{2.13}
\ene

As $e^{-it H}\var_n \to e^{-it H}\var$ in $\mathcal N(T_0)$, for $n$ large enough $\mathcal C_n$ and
$\mathcal C$ are  contractions in $\mathcal N_R(T_0)$ with the same $R,T_0,d$. The unique fixed points, $u_n,u$ are
solutions to (\ref{2.1}) in $C([0,T_0],L^2)$ with, respectively, $u_n(0)= \var_n, u(0)= \var$. Moreover, as
$\mathcal C_n(u_n)-\mathcal C (u)= \mathcal C_n(u_n)-\mathcal C_n(u)+ \mathcal C_n(u)- \mathcal C(u)$,

\beq
\|u_n-u\|_{\mathcal N(T_0)}= \| \mathcal C_n(u_n)-\mathcal C(u)\|_{\mathcal N(T_0)}
\leq   d  \| u_n-u   \|_{\mathcal N(T_0)} + \|\var_n-\var  \|_{L^2},
\label{2.14}
\ene
and as $ d < 1, u_n \to u$ in $\mathcal N(T_0)$.
As the interval of existence of the solution given in theorem \ref{th2.1} depends only on the $L^2$ norm of the initial
value, we can extend this argument, step by step, to the whole interval $[0,T]$.

\bull
\begin{remark}\label{rem2.3}
{\rm  let $T_m$ be the maximal time such that the solution $u$
given in theorem \ref{th2.1} can be extended to a solution $u \in
C([0,T_m), L^2)$ with $u(0)=\var$. Then, if $T_m$ is finite we
must have that $ \lim_{t \uparrow \infty}\|u(t)\|_{L^2}= \infty$.
In other words, the solution exists for all times unless it blows
up in the $L^2$ norm for some finite time. To prove this result
suppose that $\|u(t)\|_{L^2}$ remains bounded as $t \uparrow T_m$.
Then, by theorem \ref{th2.1} we can extend the solution
continuously to $T_m+ \epsilon$ for some $\epsilon >0$,
contradicting the definition of $T_m$.

Another consequence of theorem \ref{th2.1} is that (\ref{2.1}) has at most one solution in $C(I, L^2)$
with $u(0)=\var$. Suppose, on the contrary, that there are two, $u_1,u_2$. Then, by theorem \ref{th2.1},
$u_1(t)=u_2(t), t \in [0,T_0]$ for some $0 <T_0  < T$. Let $ T_m$ be the maximal time such that $u_1(t)=u_2(t),
t \in [0,T_m)$. Consider first the case where $ T < \infty$. Then, $T_m=T$, because if $T_m < T$, theorem \ref{th2.1}
would imply that $u_1(t)=u_2(t)$ for $ t \leq T_m+\epsilon$ for some $\epsilon > 0$, in contradiction with the definition
of $T_m$. Then, $T_m=T$ and by continuity, $u_1(T)=u_2(T)$, completing the proof in the case $T < \infty$.
A similar argument proves that if $ T=\infty$, $T_m$ can not be finite.}
\end{remark}

\bull

We will use the uniqueness  of $L^2$ solutions given by theorem \ref{2.1} and remark \ref{2.3} in the construction
of the scattering operator in theorem \ref{th3.1} in section 3.

We now study solutions in $\Q^1$.

\begin{theorem}\label{th2.4}
Suppose that assumption A holds, that $V_1$ satisfies (\ref{2.3}) and that $V_2 \in W_{1,\infty}$. Then,
for any $\var \in \Q^1$ there is a $0 < T < \infty$  such that (\ref{2.1}) has a unique solution $ u \in
C([0,T],\Q^1)$, with $u(0)= \var$. $T$  depends only on $\|\var\|_{\Q^1}$.
\end{theorem}
\noindent {\it Proof:} As $V_1$ satisfies (\ref{2.3}), it follows from (\ref{2.3b}) that,

\beq
|(V_1\var, \var)| \leq C \|\var\|_{\mathbb H^1}^2
,\quad {\rm for \, all}\, \var \in \mathbb H^1.
\label{2.15}
\ene

Then, defining $\mathcal C$ as in (\ref{2.9}) and as $V_2 \in W_{1,\infty}$,

\begin{gather}
\|\mathcal C (u)\|_{\mathcal N^1}\leq \|\var\|_{\Q^1}+ C \, T \|u\|_{\mathcal N^1}^3,
\label{2.16}
\\
\|\mathcal C (u) - \mathcal C (v) \|_{\mathcal N^1}\leq C\, T \left(\|u\|_{\mathcal N^1}^2+
 \|v\|_{\mathcal N^1}^2\right)\, \|u-v\|_{\mathcal N^1}.
 \label{2.17}
 \end{gather}

 We take $R,T$ such that
$\|\var\|_{\Q^1} + C \,T\, R^3 \leq R$ and, $d:=  C \,T\, 2\, R^2 < 1$. By (\ref{2.16}) and (\ref{2.17}), with this
choice $\mathcal C$ a contraction on $\mathcal N_R^1$ with contraction rate $d$.
The unique fixed point, $u$, is a solution to (\ref{2.1}) with $ u(0)= \var$. We prove the uniqueness of the solution
in $C([0,T],\Q^1)$ as in the proof of theorem \ref{th2.1}.

\bull

There is also continuous dependence of the solutions in $\Q^1$.

\begin{theorem}\label{th2.5}
Suppose that assumption A holds,  that $V_1$ satisfies (\ref{2.3}) and that $V_2 \in W_{1,\infty}$.
Then, the solution  $u \in C([0,T],\Q^1),0 < T < \infty,$
to (\ref{2.1})  with  $u(0)=\var$, given by theorem \ref{th2.4}, depends continuously on the initial value $\var$. In
 a precise way, let
$\var_n \to \var,$  strongly in $ \Q^1$. Then, for $n$ large enough, the solutions $u_n \in  C([0,T],\Q^1)$
to (\ref{2.1}) with initial values, $\var_n$ exist and $ u_n \to u $ in $C([0,T],\Q^1)$.
\end{theorem}

\noindent {\it Proof:} The theorem is proven as in the proof of theorem \ref{th2.2} replacing in the argument
$\mathcal N(T_0)$ by $ C([0,T_0], \Q^1)$.

\bull

\begin{remark}\label{rem2.6}{\rm
We prove as in remark \ref{rem2.3} that the solution to
(\ref{2.1}) in $\Q^1$ exists for all $ t >0$ unless it blows up in
the $\Q^1$ norm for some finite time and that theorem 2.4 implies
that (\ref{2.1}) has at most one solution on $C(I,\Q^1)$.}
\end{remark}

\bull

Let us now study solution in $\Q^2$.

\begin{theorem}\label{th2.7}
Suppose that assumption A holds, that $V_1$ satisfies (\ref{2.3}) and that $V_2 \in W_{2,\infty}$. Then,
for any $\var \in \Q^2$ there is a $0 < T < \infty$  such that (\ref{2.1}) has a unique solution $ u \in
C([0,T],\Q^2)$, with $u(0)= \var$. $T$  depends only on $\|\var\|_{\Q^2}$.
\end{theorem}

\noindent {\it Proof:} As $V_2 \in W_{2,\infty}$, for  some constant $C$,

\beq
\|V_2  \var \|_{\Q^2} \leq C \|\var\|_{\Q^2}
,\quad {\rm for \, all}\, \var \in \Q^2.
\label{2.20}
\ene

Then, with  $\mathcal C$ defined as in (\ref{2.9}),

\begin{gather}
\|\mathcal C (u)\|_{\mathcal N^2}\leq \|\var\|_{\Q^2}+ C \, T \|u\|_{\mathcal N^2}^3,
\label{2.21}
\\
\|\mathcal C (u) - \mathcal C (v) \|_{\mathcal N^2}\leq C\, T \left(\|u\|_{\mathcal N^2}^2+
 \|v\|_{\mathcal N^2}^2\right)\, \|u-v\|_{\mathcal N^2}.
 \label{2.22}
 \end{gather}

 Let $R,T$  be such that
$\|\var\|_{\Q^2} + C \,T\, R^3 \leq R$ and, $d:=  C \,T\, 2\, R^2 < 1$. Hence, it follows from
(\ref{2.21}) and (\ref{2.22})
 that $\mathcal C$ is a contraction on $\mathcal N_R^2$ with contraction rate $d$.
The unique fixed point, $u$, is a solution to (\ref{2.1}) with $ u(0)= \var$. We prove the uniqueness of the solution
in $C([0,T],\Q^2)$ as in the proof of theorem \ref{th2.1}.

\bull

\begin{theorem}\label{th2.8}
Suppose that assumption A holds,  that $V_1$ satisfies (\ref{2.3}) and that $V_2 \in W_{2,\infty}$.
Then, the solution  $u \in C([0,T],\Q^2), 0 < T < \infty,$
to (\ref{2.1})  with  $u(0)=\var$, given by theorem \ref{th2.7}, depends continuously on the initial value $\var$. In
 a precise way, let
$\var_n \to \var,$  strongly in $ \Q^2$. Then, for $n$ large enough, the solutions $u_n \in  C([0,T],\Q^2)$
to (\ref{2.1}) with initial values $\var_n$ exist and $ u_n \to u $ in $C([0,T],\Q^2)$.
\end{theorem}

\noindent {\it Proof:} The theorem is proven as in the proof of theorem \ref{th2.2} replacing in the argument
$\mathcal N(T_0)$ by $ C([0,T_0], \Q^2)$.

\bull

\begin{remark}\label{rem2.9}{\rm
We prove as in remark \ref{rem2.3} that the solution to
(\ref{2.1}) in $\Q^2$ exists for all $ t >0$ unless it blows up in
the $\Q^2$ norm for some finite time and that theorem 2.7 implies
that (\ref{2.1}) has at most one solution on $C(I,\Q^2)$.}
\end{remark}

\bull

We now consider the problem of the regularity of solutions.
Suppose that the conditions of theorems \ref{th2.1} and \ref{th2.4} are satisfied
and that $\var \in \Q^1$. Then, by theorem \ref{th2.1},
(\ref{2.1}) has a unique  $L^2$ solution and by theorem
\ref{th2.4} a unique $\Q^1$ solution, both with initial value,
$\var$. In the proposition below we prove that it is impossible
that the $\Q^1$ solution  blows-up before the $L^2$ solution.

 \begin{prop}\label{prop2.10}
 Suppose that  assumption A holds that $V_1\in L^\infty$ and that $V_2 \in W_{1,\infty}$. Let
 $ u \in C([0,T],L^2),0 <  T < \infty $ be a solution to
 (\ref{2.1}) with $u(0)=\var \in \Q^1$. Then, $u\in C([0,T],\Q^1)$.
\end{prop}

 \noindent {\it Proof:} By theorem \ref{th2.4} there is  a $ 0 < T_0 \leq T$ such that $u \in C([0,T_0],\Q^1)$.
 Let us denote, $ v:= \sqrt{H+N} u, 0\leq t \leq T_0$ . Multiplying both sides of (\ref{2.7}) by $\sqrt{H+N}$
  we obtain that,

\beq
v(t)= e^{-it H}(\sqrt{H+N})\var   + \frac{1}{i}\, G (V_1u, u)\left[(H+N)^{1/2} V_2 (H+N)^{-1/2}\right] v.
\label{2.23}
\ene
 Note that as $ V_2 \in W_{1,\infty}, \left[(H+N)^{1/2} V_2 (H+N)^{-1/2}\right]$ is a bounded operator in $L^2$.

Equation (\ref{2.23}) is a linear equation for $v$, where $u$  is
a fixed function in $\mathcal N$. Solving  this equation in an
interval $[T_0, T_0+ \Delta]$, with $\Delta $ small enough, we
prove that $v(t) \in L^2$, for  $T_0\leq t \leq T_0+\Delta$. Note
that the length of $\Delta$ depends only on $\|u\|_{ \mathcal N}$.
Repeating this argument,
 step by step, we prove that $ v \in C([0,T], L^2)$ and, in consequence, that $u \in C([0,T],\Q^1)$.

 \bull

 In the following proposition we
prove regularity between $\Q^1$ and $\Q^2$ solutions.

 \begin{prop}\label{prop2.11}
 Suppose that assumption A holds, that $V_1$ satisfies (\ref{2.19}) and that $V_2 \in W_{2,\infty}$.
 Let $ u \in C([0,T],\Q^1), 0 < T < \infty$, be a solution to
 (\ref{2.1}) with $u(0)=\var \in \Q^2$. Then, $u\in C([0,T],\Q^2)$.
\end{prop}
 \noindent {\it Proof:} By theorem \ref{th2.7} there is  a $ 0 < T_0 \leq T$ such that $u \in C([0,T_0],\Q^2)$
 and then, by (\ref{2.1}) $v:=\frac{\partial}{\partial t} u(t)\in C_B([0,T_0],L^2)$. Moreover, taking the derivative in
 time of (\ref{2.7}) we obtain that,
 \beq
 i v = e^{-itH}[H\var + F(\var)]+ G \left(  2\,\{{\rm Re}\,(v, V_1 u)\}\, V_2 u+ (V_1u,u)\, V_2 v \right).
 \label{2.24}
 \ene
Solving  the real-linear equation (\ref{2.23})- where now $u$ is a fixed function in $\mathcal N^1$- in an
interval $[T_0, T_0+ \Delta]$, with $\Delta $ small enough,
we prove that $v(t) \in L^2$, for  $T_0\leq t \leq T_0+\Delta $. Note that  as $V_1$ satisfies (\ref{2.19}),
it follows from (\ref{2.3b}) that $\|V_1 u(t)\|_{L^2}^2=(V_1^2 u(t),u(t)) \leq C \|u\|_{\mathcal N^1}^2$. In consequence,
the length of $\Delta$ depends only on $\|u\|_{ \mathcal N^1}$. Repeating this argument,
 step by step, we prove that $ v \in C([0,T], L^2)$ and, in consequence, that $u \in C([0,T],\Q^2)$.

 \bull

Let us now consider the existence of global $L^2$ solutions. For this purpose we prove that the
$L^2$ norm is constant.

\begin{lemma}\label{lem2.12}
Suppose that assumption A holds, that $V_1\in L^\infty$, that $V_2 \in W_{1,\infty}$,
and, furthermore, that $V_1,V_2$ are real valued.
Then, the $L^2$ norm of the solution to (\ref{2.1}) given by theorem \ref{th2.1} and the $L^2$ norm of the
$\Q^1$ solution given by theorem \ref{th2.4} are constant.
\end{lemma}
\noindent {\it Proof:} We first prove the theorem for the  solution $u \in C([0,T],\Q^1)$. By (\ref{2.7})
$u \in C^1([0,T], \Q^{-1})$ and then, by (\ref{2.1}),

\beq
\frac{1}{2}\frac{d}{d t}\|u(t)\|_{L^2}= \,{\rm Re}\, \left(\frac{d}{d t}u(t),u(t)\right)= \,{\rm Re}\, \frac{1}{i}
\left[( Hu,u)+(V_1u,u)\, (V_2u,u) \right]=0,
\label{2.25}
\ene
and then, $\|u(t)\|_{L^2}= \|u(0)\|_{L^2}$. The result in the case of solutions $u  \in C([0,T],L^2)$
follows approximating $u(0)=\var$ in the $L^2$ norm by $ \var_n \in \Q^1$ and applying theorem \ref{th2.2}.

\bull

\begin{theorem}\label{th2.13}
Suppose that assumption A holds, that $V_1\in L^\infty$, that $V_2 \in W_{1,\infty}$,
and, furthermore, that $V_1,V_2$ are real valued. Then, the $L^2$ solution, $u(t)$, given by theorem \ref{2.1}
exists for all times, and $\|u(t)\|_{L^2}= \|u(0)\|_{L^2}, t \in [0,\infty)$.
\end{theorem}
\noindent {\it Proof:} the theorem follows from remark \ref{rem2.3} and lemma \ref{lem2.12}.

\bull

For $ u(t) \in C(I,\Q^1)$ we define the energy at time t  as
follows, \beq E(u(t)):=(u'(t),u'(t))+(V_0 u(t), u(t))+
\frac{1}{2}(V_1 u(t), u(t))\,(V_2u(t),u(t)). \label{2.26} \ene
\begin{lemma}\label{lem2.14}
Suppose that assumption A holds, that $V_1= \lambda V_2$, for some
real $\lambda$, that $V_2$ is real valued, and that $V_2 \in
W_{2,\infty}$. Then, the energy of the $\Q^1$ solution to
(\ref{2.1}) given by theorem \ref{th2.4} and the energy of the
$\Q^2$ solution to (\ref{2.1}) given by theorem \ref{th2.7} are
constant in time.
\end{lemma}
\noindent {\it Proof:} We first prove the lemma for  solutions $u
\in C([0,T], \Q^2)$. If follows from (\ref{2.7}) that, $u \in
C^1([0,T], L^2)$. As $ \Q^2 =D(H)$, we can write the energy as
follows,

$$
E(u(t)):=( u(t),H u(t))+ \frac{\lambda}{2}(V_2 u(t), u(t))^2.
$$
It follows that,
\begin{gather*}
\frac{d}{dt} E(u(t))= 2 \,{\rm Re}\, \left[ (\dot{u}(t), H u(t))+
\lambda (V_2 u(t),u(t)) (V_2 \dot{u}(t),u(t))\right]=
\\
=  2 \,{\rm Re}\, \frac{1}{i} \left[\|Hu(t)\|_{L^2}^2+ \lambda (V_2
u(t), u(t))\,\{ (V_2 u(t),Hu(t)) +(Hu(t), V_2 u(t))\}\right]=0.
\end{gather*}

The result in the case of solutions $u  \in C([0,T],\Q^1)$ follows
approximating $u(0)=\var$ in the $\Q^1$ norm by $ \var_n \in \Q^2$
and applying theorem \ref{th2.5}.

\bull
\begin{theorem}\label{th2.15}
Suppose that assumption A holds, that $V_1= \lambda V_2$, for some
real $\lambda$, that $V_2$ is real valued, and that $V_2 \in
W_{2,\infty}$. Then,  the $\Q^1$ solution to (\ref{2.1}) given by
theorem \ref{th2.4} and  the $\Q^2$ solution to (\ref{2.1}) given by
theorem \ref{th2.7} exist for all times and the $L^2$ norm and the
energy of the solutions are constant in time.
\end{theorem}
\noindent {\it Proof:} Let us first consider  the solution $ u \in
C(I, \Q^1)$. By lemmata \ref{lem2.12} and \ref{lem2.14}
$$
(u'(t),u'(t))+(V_0 u(t), u(t))+N (u(t),u(t))\leq E(u(0))+ N\,
\|u(0)\|^2_{L^2}+ \frac{|\lambda|}{2} \|V_2\|^2_{L^\infty}
\|u(0)\|_{L^2}^4.
$$
Then, by remark \ref{rem2.6} $u$ exists for all times, and the $L^2$
norm and the energy are constant in time. The theorem follows in the
case of $\Q^2$ solutions by proposition \ref{prop2.11}.

\section{Scattering}\sss
In this section we construct the small amplitude scattering
operator, $S,$ for equation (\ref{1.2}) and we give a method for the unique
reconstruction of the potential $V_0$ and the coupling constant
$\lambda$, from $S$.

We first introduce some standard notations and some results that
we need.

For any $\gamma \in \ere$ let us denote by $L^2_{\gamma}$ the
Banach space of all complex-valued measurable functions on $\ere$
such that,

\beq \left\|\varphi \right\|_{\ds L^1_{\gamma}} := \int |\varphi(x)|\, (1+|x|)^\gamma\, dx < \infty.
\label{3.1}
\ene

If $ V_0 \in L^1_1$ the differential expression $\tau:= -\frac{d^2}{dx^2}+V_0(x)$ is essentially self-adjoint
on the domain,

$$
D(\tau):= \left\{\varphi \in L^2_C: \varphi,\, {\rm and} \,  \varphi' \, \hbox{\rm are absolutely
continuous and}
\, \tau \varphi \in L^2 \right\},
$$
where $L^2_C$ denotes the set of all functions in $L^2$ that have compact support. We denote by $H$ the
unique self-adjoint realization of $\tau$. As is well known, \cite{dt}, \cite{wei}, $H$ has a finite number of
negative eigenvalues, it has no positive of zero eigenvalues, it has no singular-continuous spectrum, and the
absolutely-continuous spectrum is $[0,\infty)$. By $H_0$ we denote the unique self-adjoint realization of
$-\frac{d^2}{d x^2}$ with domain $\mathbb H ^2$. The wave operators are defined as follows,
$$
W_{\pm}:= {\rm s-} \lim_{t\rightarrow \pm \infty}\, e^{itH} \, e^{-it H_0}.
$$
The limits  above exist in the strong topology in $L^2$ and  ${\rm Range}\, W_{\pm} = \mathcal H_{ac}$,
where $\mathcal H_{ac}$ denotes the space of absolute continuity of $H$. Moreover, the intertwining relations
hold, $H W_{\pm}= W_{\pm} H_0$. for these results see \cite{sch}. The linear scattering operator is defined as

\beq
S_L:= W_+^\ast \, W_-.
\label{3.2}
\ene

For any pair, $\varphi, \psi,$ of solutions to the stationary Schr\"odinger equation

\beq
-\frac{d^2}{dx^2}\varphi +V_0 \varphi =k^2 \varphi, k \in \CE,
\label{3.3}
\ene
let $[\varphi,\psi]$ denote the Wronskian of $\varphi$ and $\psi$,

$$
[\varphi,\psi]:= \varphi' \psi-\varphi \psi'.
$$

Let $f_j(x,k), j=1,2$ be the Jost solutions to (\ref{3.3}) that satisfy, $ f_1(x,k)\approx e^{ik x}, x \to \infty,
f_2(x,k)\approx e^{-ikx}, x \to - \infty$, \cite{fa1,fa2,dt,cs}. The potential $V_0$ is said to be {\it generic} if
$[f_1(x,0), f_2(x,0)]\neq 0$, and it is said to be {\it exceptional} if $[f_1(x,0), f_2(x,0)]= 0$. When $V_0$ is
{\it exceptional} there is a bounded solution to (\ref{3.3}) with $k^2 =0$, that is called a half-bound state
or a zero energy resonance. The trivial potential $V_0=0$ is exceptional.

Below we will always assume that $V_0 \in L^1_{\gamma}$, where in
the {\it generic case} $\gamma > 3/2$ and in the {\it exceptional
case} $\gamma > 5/2$.

In Theorem 1.1 of \cite{we1} it was proven  that the operators
$W_{\pm}$ and $W^{\ast}_{\pm}$ are bounded on $W_{j,p}, j=0,1, 1
< p < \infty$.

By Theorem 3 in page 135 of \cite{st}

\beq \|{\mathcal F}^{-1} (1+q^2)^{j/2}({\mathcal F} f)(q)\|_{L^p},
\label{3.4}
\ene
is a norm that is equivalent to the norm of
$W_{j,p}, 1 < p < \infty$. In (\ref{3.4}) ${\mathcal F}$ denotes the
Fourier transform.

If $H$ has no eigenvalues the $W_{\pm}$ are unitary operators on $L^2$, and it follows from the intertwining
relations  that

$$
(1+H)^{j/2}= W_{\pm}\, (1+H_0)^{j/2}\,W_{\pm}^{\ast},
$$
and then, by (\ref{3.4}),
 \beq
 \|(I+H)^{j/2}\,f \|_{L^p},
 \label{3.5}
 \ene
 defines a norm that is equivalent to the norm of $W_{j,p}, j=0,1, 1 < p < \infty$. Below
we  use this equivalence without further comments. Furthermore, the following $L^p-L^{p'}$ estimate holds

\beq
\left\|e^{-it H}\right\|_{\ds{\mathcal B}\left( W_{1,p} ,W_{1,p'}\right)}\, \leq C\frac{1}{\ds
|t|^{\frac{1}{p}-\frac{1}{2}}},
1 \leq p \leq 2,
\label{3.6}
\ene
 where for any pair of Banach spaces $\mathcal X, \mathcal Y, \mathcal B(\mathcal X,\mathcal Y)$ denotes the Banach
 space of all bounded operators from $\mathcal X$ into $ \mathcal Y$. When $H$ has bound states
 estimate (\ref{3.6}) is proven in \cite{we2} for the restriction of $e^{-it H}$ to the
subspace of continuity of $H$.

The norm (\ref{3.5}) for the Sobolev spaces $W_{j,p}, 1 < p < \infty$, and the $L^p-L^{p'}$ estimate (\ref{3.6})
are the basic tools that we use in order to construct the scattering operator and to solve the inverse scattering
problem.

For any $1 < q < 3/2$ we denote, $p:= (q+1)/(q-1),$  $r:=
(4q)/(2q-1)$. We designate, $P:=\left(1/r, 1/(p+1)\right)$ and we define,

$$
L(P):= L^r\left(\ere, L^{p+1}\right).
$$

Let $\mathcal M$ be the following Banach space,
\beq
\mathcal M:= C_B\left(\ere, L^{p+1}\right) \cap
L(P),
\label{3.7}
\ene
with norm,

$$
\|\varphi\|_{\mathcal M}:= {\mathrm max }\left[\|\varphi\|_{ C_B\left(\ere,\, L^{p+1} \right)}
, \|\varphi\|_{L(P)}  \right].
$$
Recall that $C_B\left(\ere, L^{p+1}\right)$ denotes the Banach space of all bounded and continuous
functions from $\ere$
into $L^{p+1}$.

 In the following theorem we construct the small amplitude
non-linear scattering operator.
\begin{theorem}
\label{th3.1}
Suppose that $V_0 \in L^1_{\gamma}$ where in the {\it generic
case} $\gamma > 3/2$ and in the {\it exceptional case} $ \gamma >
5/2$ and that $H$ has no eigenvalues. Moreover, assume that $V_j
\in L^q \cap  L^\infty$ for some $1 < q < 3/2$. Then, there is a $\delta > 0$
such that for every $ \varphi_-\in \mathbb H^1$ with
$\|\varphi_-\|_{\ds \mathbb H^1} < \delta$ there is a unique
solution, $u,$ to (\ref{1.2}) such that $u\in \mathcal M \cap
C_B(\ere,L^2)$ and
\beq \lim_{t\to -\infty}\|u(t)-e^{-it
H}\varphi_-\|_{\ds L^2}=0.
\label{3.8}
\ene
Moreover, there is a
unique $\varphi_+ \in L^2$ such that

\beq
\lim_{t\to \infty}\|u(t)-e^{-it H}\varphi_+\|_{\ds L^2}=0.
\label{3.9}
\ene
Furthermore, $e^{-itH}\varphi_{\pm} \in \mathcal M$ and

\beq
\left\| u(t)- e^{-itH} \phi_{\pm}\right\|_{\mathcal M}\leq C \left\| e^{-it H} \varphi_{\pm}
\right\|^{3}_{\mathcal M},
\label{3.10}
\ene

\beq
\left\|\varphi_+ -\varphi_- \right\|_{\ds L^2} \leq C \left\|\varphi_-\right\|_{\ds \mathbb H^1}^3.
\label{3.11}
\ene
The scattering operator $S_{ V_0}:\varphi_- \hookrightarrow \varphi_+$ is injective.
\end{theorem}

\noindent{\it Proof:} Observe that $u \in C_B(\ere,
 L^2)\cap\mathcal M$ is a solution to (\ref{1.2}) with $\lim_{t\to -\infty}\|u(t)-e^{-it H_0}
 \varphi_-\|_{\ds L^2}=0$ if and only if it is a solution to the
 following integral equation (this is proven as in the proof of the equivalence of (\ref{2.1}) and (\ref{2.7}))

\beq
u= e^{-itH} \varphi_- +\frac{1}{i} \int_{-\infty}^{t}
e^{-i(t-\tau)\,H} F(u(\tau))\, d \tau,
\label{3.12}
\ene
where,

\beq
F(u):= \lambda \left( V_1 u,u \right) \, V_2 \, u.
\label{3.13}
\ene
We will prove that the integral in the right-hand side of (\ref{3.12}) is absolutely convergent in $\mathcal M$ and in
$L^2$.

For $u\in \mathcal M$ we define,

\beq
\mathcal P (u)(t):= \int_{-\infty}^t e^{-i(t-\tau)H}\, F(u(\tau))\, d \tau.
\label{3.14}
\ene

By (\ref{3.6}) and H\"older's inequality,
\beq
\begin{array}{ll}
\|( \mathcal Pu)(t)\|_{\1}\leq C \int_{-\infty}^t\,  \frac{1}{|t-\tau|^d}\,\|F(u(\tau))\|_{\p} \\ \\
\leq C \,\int_{-\infty}^t\,\frac{1}{|t-\tau|^d}\, \|u(\tau)\|^3_{\1}\, d\tau,
\\
\end{array}
\label{3.15}
\ene
where, $d:=1/(2q)$. Then,
\beq
\|( \mathcal Pu)(t)\|_{\1} \leq C \, (I_1+I_2),
\label{3.16}
\ene
where,
$$
I_1:= \int_{-\infty}^{t-1}\,\frac{1}{|t-\tau|^d}\, \|u(\tau)\|^3_{\1}\, d\tau,
$$
and
$$
I_2:= \int_{t-1}^{t}\,\frac{1}{|t-\tau|^d}\, \|u(\tau)\|^3_{\1}\, d\tau.
$$

Let us denote by $\chi_{(1,\infty)}$ the characteristic function  of $(1,\infty)$. Then, by H\"older's inequality,
$$
I_1 \leq \left\|\chi_{(1,\infty)}(\tau) \frac{1}{|\tau|^d}\right\|_{L^{\alpha}}\,\|u\|_{L(P)}^3 ,
$$
where, $\alpha:= r/(r-3)$. Note that as $d \alpha >1,\,
 \|\chi_{(1,\infty)}(\tau) \frac{1}{|\tau|^d}\|_{L^{\alpha}} < \infty$.

Moreover, as $ d < 1$,
$$
I_2 \leq C \|u\|_{C_B(\ere, \1  )}^3.
$$
It follows that,
\beq
\|( \mathcal Pu)(t)\|_{\1} \leq C \,\|u\|_{\mathcal M}^3.
\label{3.17}
\ene
We prove in a similar way that $( \mathcal Pu)(\cdot)$ is a continuous function on $\ere$ with values in
$\1$.

Furthermore, it follows from (\ref{3.15})
 and the generalized Young inequality \cite{rs2}, that,
\beq
\|\mathcal P(u)\|_{\P}\leq C  \|u\|_{\P}^3.
\label{3.18}
\ene

Then, by (\ref{3.17}) and (\ref{3.18}),

\beq
\| \mathcal Pu\|_{\mathcal M} \leq C \,\|u\|_{\mathcal M}^3.
\label{3.19}
\ene

In an analogous way we prove that,

\beq \| \mathcal P u- \mathcal P v\|_{\mathcal M} \leq C
\, (\|u\|_{\mathcal M}^2+ \|v\|_{\mathcal
M}^2)\,  \|u-v\|_{\mathcal M}.
\label{3.20}
\ene
Furthermore,
\begin{gather*}
\left\| (\mathcal P u)(t)\right\|_{L^2}^2= \int_{-\infty}^t\, d\tau \,\left( F(u(\tau)),
\int_{-\infty}^t\, e^{-i (\tau-\tau')H}\, F(u(\tau'))\right)= \\
2 {\mathrm Re}\int_{-\infty}^t\left( F(u(\tau)), \mathcal P(u)(\tau)\right)\, d\tau.
\end{gather*}
We define,
$$
u_t(\tau):= \chi_{(-\infty, t)}(\tau)\, u(\tau), \qquad g_t(\tau):=
\int_{-\infty}^{\tau} \frac{1}{|\tau -\tau'|^d} \|u_t(\tau')\|^3_{\1}\, d\tau.
$$
Then, by (\ref{3.6}) and H\"older's inequality,

\begin{gather}
\left\| (\mathcal P u)(t)\right\|_{L^2}^2 \leq  \int_{-\infty}^t
\|u_t(\tau)\|_{L^{p+1}}^3\, g_t(\tau)\, d\tau, \nonumber\\
\leq C \, \left[\int \, \|u_t(\tau)\|^{3r/(r-1)}_{\1}\,d\tau\,\right]^{(r-1)/r}
\, \|g_t\|_{L^r}.
\label{3.21}
\end{gather}
By the generalized Young's inequality,

$$
\|g_t\|_{L^r}\leq C \|u_t\|_{\P}^3.
$$
Furthermore,
$$
\left[\int \, \|u_t(\tau)\|^{3r/(r-1)}_{\1}\,d\tau\,\right]^{(r-1)/r} \leq \|u\|_{\m}^{4-r}\,\|u_t\|_{\P}^{r-1}.
$$
Hence, by (\ref{3.21})
\beq
\left\| (\mathcal P u)(t)\right\|_{L^2}^2 \leq \, \|u\|_\m^{4-r} \|u_t\|_{\m}^{r+2}.
\label{3.22}
\ene
We prove in a similar way that the function $ t \in \ere \to  (\mathcal P u)(t)$ with values in $L^2$ is continuous.

We first prove that equation (\ref{3.12}) has at most one solution in $\m$ and then, we prove the existence of a
solution for $\varphi$ small.

Suppose that there are two solutions  in $\m$ to (\ref{3.12}), $u,v,$ and denote, $u_T:= \chi_{(-\infty, T)} u,v_T:=
\chi_{(-\infty,T)} v$. Then,

\beq
u_T(t)-v_T(t)= (\mathcal P u_T)(t)-(\mathcal P v_T)(t), \,{\mathrm for}\, t\leq T.
\label{3.23}
\ene
Arguing as in the proof of (\ref{3.18}), and as $u_T(t)=v_T(t)=0$ for $t \geq T,$ we prove that,

\beq
\| u_T- v_T\|_{\P} \leq C \,
(\|u_T\|_{\P}^2+ \|v_T\|_{\P}^2)\,  \|u_T-v_T\|_{\P}.
\label{3.24}
\ene
As $\lim_{T\to -\infty}\,(\|u_T\|_{\P}^2+ \|v\|_{\P}^2)=0$ we can take $T$ so negative that,
$$
C \,
(\|u_T\|_{\P}^2+ \|v\|_{\P}^2) < 1/2,
$$
where $C$ is the constant in (\ref{3.24}). Then, for such $T$ equation (\ref{3.24}) implies that,

$$
\| u_T-  v_T\|_{\P} < \frac{1}{2}\, \| u_T-   v_T\|_{\P},
$$
and then, $u_T(t)=v_T(t), t \leq T$ and by the uniqueness of the initial value problem at finite time
(see theorem \ref{th2.1}),
$u(t)=v(t), t \in \ere$.

We now prove that $e^{-itH}\in {\mathcal B}\left(\mathbb H^1, \m \right)$.

Since $e^{-itH}$ is a strongly
continuous unitary group on $L^2$ that commutes with $(1+H)^{1/2}$ we have that, $ e^{-itH} \in
{\mathcal B}\left(\mathbb H^1,C_B (\ere, \mathbb H^1 )
\right)$. Furthermore, as by Sobolev's  theorem \cite{ad} and interpolation \cite{rs2}
$\mathbb H^1$ is continuously imbedded in $\1$, it follows that, $e^{-itH} \in
{\mathcal B}\left(\mathbb H^1, C_B(\ere, \1) \right).$

Moreover, by (\ref{3.6}) and Lemma 3.1 of \cite{ka2} it follows that (in \cite{ka2} the operator $e^{-itH_0}$ is
considered, but the same proof applies in our case)  $ e^{-itH} \in
{\mathcal B}\left( L^2,L^r (\ere, L^{l} )\right)$ with, $l=2r/(r-2)$, and then, we have that,
$e^{-itH}\in{\mathcal B}\left(\mathbb H^1,L^r (\ere, W_{1,l} )\right)$. Furthermore, as by Sobolev's theorem and
interpolation \cite{rs2}, $W_{1,l}$ is
continuously imbedded in $\1, $ we have that,
 $e^{-itH}\in{\mathcal B}\left(\mathbb H^1,L^r (\ere, \1 )\right).$
 Then,

\beq
e^{-itH}\in {\mathcal B}\left(\mathbb H^1, \m \right).
\label{3.25}
\ene

As in section 2, for $ R >0$  we denote,

$$
\m_R:=\{ u \in \m : \| u\|_\m \leq R\}.
$$
We now take $R$ so small that $  C \max [R^3, (2R)^2] < 1/2$, where $C$ is the
biggest of the constants  in (\ref{3.19}) and (\ref{3.20}), and $\delta$ so small
that
$$
\|e^{-itH}\varphi\|_\m \leq R/4 ,\, {\mathrm for}\, \|\varphi\|_{\mathbb H^1 }< \delta.
$$
Then if $\|\varphi\|_{\mathbb H^1 }< \delta$ the operator,
\beq
\mathcal C(u):= e^{-itH}\varphi+ \mathcal P(u)
\label{3.26}
\ene
is a contraction on $\m_R$. By the contraction mapping theorem \cite{rs1} $\mathcal C$ has a unique fixed point
in $\m_R$ that is a solution to (\ref{3.12}), and moreover,

\beq
\|u\|_\m \leq \|e^{-itH}\varphi\|_\m + \frac{1}{2} \|u\|_\m,
\label{3.27}
\ene
and hence,
\beq
\|u\|_\m \leq  2   \|e^{-itH}\varphi_-\|_\m.
\label{3.28}
\ene
By (\ref{3.12}) and (\ref{3.22}), $u \in C_B(\ere, L^2)$ and (\ref{3.8}) holds.
 Equations (\ref{3.12}), (\ref{3.19}) and (\ref{3.28}) imply that (\ref{3.10}) holds for $\varphi_-$.
We define,

\beq
\varphi_+ := \varphi_-+ \frac{1}{i}\, \int_{-\infty}^{\infty}
e^{i\tau\,H} F(u(\tau))\, d \tau.
\label{3.29}
\ene
Estimating as in the proof of (\ref{3.17}) we prove that $\varphi_+ \in \1$, and arguing as in the proof of
(\ref{3.22}) it follows that $\varphi_+ \in L^2$ and that
$$
\|\varphi_+-\varphi_-\|_{L^2}\leq C \|u\|_\m^3 \leq C \|e^{-itH}\varphi\|_\m^3.
$$
Equation (\ref{3.11}) follows now from (\ref{3.25}).

By (\ref{3.12}) and (\ref{3.29}),

\beq
u= e^{-itH} \varphi_+ -\frac{1}{i} \int_{t}^{\infty}
e^{-i(t-\tau)\,H} F(u(\tau))\, d \tau.
\label{3.30}
\ene
Equation (\ref{3.9}) follows from (\ref{3.30}) estimating as in the proof of (\ref{3.22}).
Moreover, estimating as in the proof of (\ref{3.19}) we have that,
\beq
\left\| \int_{t}^{\infty}
e^{-i(t-\tau)\,H} F(u(\tau))\, d \tau \right\|_\m \leq C \|u\|_\m^3.
\label{3.31}
\ene

Multiplying both sides of (\ref{3.29}) by $e^{-itH}$ and estimating as in the proof of (\ref{3.19}) we prove that
$e^{-itH}\varphi_+ \in \m$. By (\ref{3.30}), (\ref{3.31}) and arguing as in the proof of (\ref{3.28}) we obtain that,
\beq
\|u\|_\m \leq  2   \|e^{-itH}\varphi_+\|_\m.
\label{3.32}
\ene
At this point, (\ref{3.10}) for $\varphi_+$ follows from (\ref{3.30})--(\ref{3.32}). Note that the uniqueness of
 $\varphi_+$ is immediate from the fact that  $e^{-itH}$ is unitary on $L^2$.

 Finally, we prove that $S_{V_0}$ is injective. Suppose that $\varphi_+= S_{V_0}\varphi_-=0$. Then, by (\ref{3.30})
\beq
u=  -\frac{1}{i} \int_{t}^{\infty}
e^{-i(t-\tau)\,H} F(u(\tau))\, d \tau.
\label{3.33}
\ene
We prove that (\ref{3.33}) implies that $u=0$ arguing as in the proof of the uniqueness of the solution to equation
(\ref{3.12}) and then, it follows from (\ref{3.8}) that $\varphi_-=0$.

\bull

We now define the scattering operator that relates  asymptotic states that are solutions to the free Schr\"odinger
equation,

$$
i\frac{\partial}{\partial t}u= H_0 u,
$$
given by,
\beq
S:= W_+^\ast\, S_{\ds V_0}\, W_-.
\label{3.34}
\ene
In the following theorem we show that we can uniquely reconstruct the linear scattering operator from the small
amplitude behaviour of $S$.

\begin{theorem}
\label{th3.2}
Suppose that the assumptions of Theorem \ref{th3.1} are satisfied. Then, for every $\varphi \in \mathbb H^1$,
\beq
\frac{d}{d\epsilon} S(\epsilon \varphi)= S_L\, \varphi,
\label{3.35}
\ene
where the derivative exists in the strong convergence in $L^2$.
\end{theorem}

\noindent{\it Proof:} Since $S(0)=0$ and the wave operators $W_{\pm}$  are bounded on $\mathbb H^1$ \cite{we1}
 it is sufficient to prove that,
\beq
{\mathrm s-}\lim_{\epsilon \to 0}\frac{1}{\epsilon}\left[ S_{\ds V_0}(\epsilon \varphi)-\epsilon \varphi
\right]=0.
\label{3.36}
\ene
But (\ref{3.36}) follows from (\ref{3.11}) with $\varphi_-$ replaced by $\epsilon \varphi$.

\bull
\begin{corollary}
\label{cor3.3}
Suppose that the assumptions of Theorem \ref{th3.1} are satisfied. Then, $S$ uniquely determines the linear potential
$V_0$.
\end{corollary}
\noindent{\it Proof:} By Theorem \ref{th3.2} we uniquely
reconstruct $S_L$ from $S$. From $S_L$ we obtain the reflection
coefficients for linear Schr\"odinger scattering on the line (see
Section 9.7 of \cite{pe}). As $H$ has no bound states we uniquely
reconstruct $V_0$ from one of the reflection coefficients using
any of the standard methods. See, for example, \cite{fa1,fa2},
\cite{dt}, \cite{me}, \cite{mar}, \cite{cs}, \cite{ak}.

\bull

Note that our proof gives a constructive method to uniquely reconstruct $V_0$ from $S$. We first compute $S_L$
from the derivative in (\ref{3.35})and then, we obtain the reflection coefficients and we reconstruct $V_0$ from one of
them.

The following Theorem gives us a convergent expansion at low amplitude for $S_{\ds V_0}$

\begin{theorem}
\label{th3.4}
Suppose that the assumptions of Theorem \ref{th3.1} are satisfied. Then, for any $\varphi \in \mathbb H^1$,
\begin{gather}
i\left((S_{\ds V_0}-I)(\epsilon \varphi),\varphi\right)= \epsilon^3 \lambda \int_{-\infty}^{\infty}\,
\left(V_1e^{-itH}\varphi,e^{-itH}\varphi\right)\, \left(V_2e^{-itH}\varphi,e^{-itH}\varphi\right)\, dt
\nonumber \\
+ O(\epsilon^5),
\quad \epsilon \to 0.
\label{3.40}
\end{gather}
\end{theorem}

\noindent{\it Proof:} Suppose that $u_j \in \m , j=1,\cdots, 4$. Then, by H\"older's inequality,

\begin{gather}
\left| \int \,(V_1 u_1(t),u_2(t))\, (V_2 u_3(t),u_4(t))\, dt\right|  \nonumber \\
\leq  \|V_1\|_{L^q}\, \|V_2\|_{L^q}\, \int \|u_1(t)\|_{\1} \|u_2(t)\|_{\1} \|u_3(t)\|_{\1} \|u_4(t)\|_{\1}\, dt
\nonumber
\\
\leq  \|V_1\|_{L^q}\, \|V_2\|_{L^q}\,\|u_1\|_{\P}\, \|u_2\|_{\P}\,
\, \|u_3\|_{\P}\, \left[\int \,
 \|u_4(t)\|^{r/(r-3)}_{\1}\, dt\right]^{(r-3)/r}\nonumber \\
 \leq \|V_1\|_{L^q}\, \|V_2\|_{L^q}\,\|u_1\|_{\P}\, \|u_2\|_{\P}\, \, \|u_3\|_{\P}
\, \|u_4\|^{4-r}_{\m}\,\|u_4\|_{\P}^{r-3}\nonumber \\
\leq   \|V_1\|_{L^q}\, \|V_2\|_{L^q} \Pi_{j=1}^4\,
 \|u_j\|_{\m}.
\label{3.41}
\end{gather}

 As $e^{-itH}\varphi \in \m$, (\ref{3.41}) with $u_j= e^{-itH}\varphi, j=1,2,3,4$, proves that the integral in
 the right-hand side of (\ref{3.40}) is absolutely
 convergent.

By the contraction mapping theorem, the solution $u$ that satisfies (\ref{3.8}) with $ \epsilon \varphi$
instead of $\varphi_-$ is given by,
\beq
u(t)= \lim _{j \to \infty}{\mathcal C}^j (\epsilon \,e^{-itH}\varphi) =\epsilon \,e^{-itH}\varphi  + v(t), \,{\rm where}\,
v(t)= \sum_{j=1}^\infty {\mathcal P}^j(\epsilon e^{-itH}\varphi).
\label{3.42}
\ene
Moreover,  it follows from (\ref{3.19}) that if $\epsilon$ is small enough,
\beq
\|v\|_{\m} \leq C\, \epsilon^3.
\label{3.43}
\ene
Then,  (\ref{3.40}) follows by (\ref{3.29}) with $\epsilon \varphi$ instead of $\varphi_-$ and (\ref{3.41})-(\ref{3.43}).

\bull
\begin{corollary}\label{cor3.5}
Suppose that the conditions of Theorem \ref{th3.1} are satisfied and that $V_1,V_2,$ are  real-valued functions that
are not identically zero. Moreover, assume  either that
$V_1= V_2$ or that  $V_j, j=1,2$, do not change sign and  $V_1\, V_2 \ne 0$ in a set of positive measure. Then, the
scattering operator, $S$, and $V_j,j=1,2$, determine uniquely
$\lambda$.
\end{corollary}

\noindent {\it Proof:}

By (\ref{3.40})
\beq
\lambda = \lim_{\epsilon \to 0}\frac{1}{\epsilon^3} \frac{i\left((S_{\ds V_0}-I)(\epsilon \varphi),\varphi\right)}
{\int_{-\infty}^{\infty}\,
\left(V_1e^{-itH}\varphi,e^{-itH}\varphi\right)\, \left(V_2e^{-itH}\varphi,e^{-itH}\varphi\right)\, dt}.
\label{3.44}
\ene
 By corollary \ref{cor3.3} $V_0$ is known, and then $H:= H_0+V_0$ is known. Then, $W_{\pm}$ are known,
 and $S$ uniquely determines $S_{\ds V_0}$ (see (\ref{3.34})). Hence, the right-hand side of (\ref{3.44})
 is uniquely determined by our data. Moreover, under our conditions we can always find a $\varphi \in \mathbb H^1$
 such that the denominator of the right-hand side of(\ref{3.44}) is not zero.

\bull

Note that (\ref{3.44}) gives us a formula for the reconstruction of $\lambda$.

Let us now go back to the quantum capacitor (\ref{1.1}) where we take a slightly more general external potential $V_0$,
namely,

$$
V_0(x)=  \left[\beta_1 \chi_{[a,b]}(x)+ \beta_2\chi_{[c,d]}(x)\right],
$$
where $\beta_1,\beta_2 \in \ere$. By corollary \ref{cor3.3} we uniquely reconstruct $V_0$ from  $S$. Then,
$\beta_1,\beta_2, a,b,c$ and $d$ are uniquely reconstructed. Moreover, by corollary \ref{cor3.5} we uniquely reconstruct
$\lambda$. Hence, from $S$ we uniquely reconstruct all the physical parameters of the quantum capacitor.


\end{document}